\documentclass[11pt,letter]{article}

\usepackage{amssymb,amsmath,url}
\usepackage{graphicx,color}

\usepackage[margin=1.0in]{geometry}

\newcommand{\RR}{{\bf R}}
\newcommand{\RS}{{\bf S}}

\begin{document}

\title{\bf An Efficient Algorithm for Upper Bound on the Partition Function of Nucleic Acids}

\author{Hamidreza Chitsaz\\ 
Department of Computer Science \\
Wayne State University  \\
Detroit, MI 48202 \\
\url{chitsaz@wayne.edu}
\and Elmirasadat Forouzmand \\
Department of Computer Science \\
Wayne State University  \\
Detroit, MI 48202 \\
\url{elmira@wayne.edu}
\and Gholamreza Haffari \\
Faculty of Information Technology\\
Monash University\\
Clayton, VIC 3800 \\
Australia \\ 
\url{reza@monash.edu}
}

\maketitle

\begin{abstract}
It has been shown that minimum free energy structure for RNAs and RNA-RNA interaction is often incorrect due to inaccuracies in the energy parameters and inherent limitations of the energy model. In contrast, ensemble based quantities such as melting temperature and equilibrium concentrations can be more reliably predicted. Even structure prediction by sampling from the ensemble and clustering those structures by Sfold \cite{DinLaw03} has proven to be more reliable than minimum free energy structure prediction.
The main obstacle for ensemble based approaches is the computational complexity of the partition function and base pairing probabilities. For instance, the space complexity of the partition function for RNA-RNA interaction is $O(n^4)$ and the time complexity is $O(n^6)$ which are prohibitively large \cite{Chitsaz09,Huang09}. Our goal in this paper is to give a fast algorithm, based on sparse folding, to calculate an upper bound on the partition function. Our work is based on the recent algorithm of Hazan and Jaakkola \cite{Hazan12}. The space complexity of our algorithm is the same as that of sparse folding algorithms, and the time complexity of our algorithm is $O(MFE(n)\ell)$ for single RNA and $O(MFE(m, n)\ell)$ for RNA-RNA interaction in practice, in which $MFE$ is the running time of sparse folding and $\ell \leq n$ ($\ell \leq n + m$) is a sequence dependent parameter.

\end{abstract}

\section{Introduction}
Since the turn of the millennium and the advent of high throughput biology in the post-genome era, startling discoveries have redefined the role of RNA as a key player in the cellular arena. The ribosome and spliceosome are essentially two major RNA machines that together with other structural RNAs such as microRNAs (miRNA), long intergenic non-coding RNAs (lincRNA), small bacterial non-coding RNAs, and many other categories of structural RNAs run the cell at an extent that is comparable to that of protein machinery. 
For instance, lincRNAs have been recently shown to play sophisticated regulatory roles in mammalian cells, and miRNAs play a significant role in the development of cancer. These discoveries have put RNA together with proteins in the center of focus for research and therapeutic purposes, including personalized medicine. Humanity has just begun to unravel RNA's complicated roles in living cells, and RNA is no longer considered a mere information medium from DNA to proteins, but it is rather in the center of attention in molecular and cellular biology research. RNA molecules often function through interaction with other RNAs.
In the absence of high throughput experimental assays to observe RNA structure and RNA-RNA interactions,  the problems of RNA structure prediction and RNA-RNA interaction prediction gain the highest priority in bioinformatics.

RNA structure and RNA-RNA interaction prediction have recently received significant attention. The majority of algorithms that have been developed predict the minimum free energy structure \cite{AlkKarNadSahZha06} or binding sites \cite{Muckstein08}. However, it is a well-known fact that minimum free energy structure is often incorrect due to inaccuracies in the energy parameters and inherent limitations of the energy model. On the other hand, it has been shown that thermodynamic quantities, such as melting temperature and equilibrium concentrations, that are derived from the partition function which captures the properties of the whole Boltzmann ensemble rather than those of the most likely structure, can be more reliably predicted \cite{Chitsaz09}. Even structure prediction by sampling from the ensemble and clustering those structures by Sfold \cite{DinLaw03} has proven to be more reliable than minimum free energy structure prediction \cite{Ding05,Ding06}.

The main obstacle for ensemble based approaches is the computational complexity of the partition function and base pairing probabilities. For instance, the space complexity of computing the partition function for RNA-RNA interaction is $O(n^4)$ and the  time complexity is $O(n^6)$ which are prohibitively large \cite{Chitsaz09,Huang09}. On the other hand, recent progress in sparse folding algorithms has provided fast algorithms for the prediction of the most likely (minimum free energy) structure \cite{Backofen11,Mohl10,Salari10}. Although the partition function cannot be calculated exactly using sparsification ideas, it may be approximated. Our goal in this paper is to give a fast algorithm, based on sparse folding, to calculate an upper bound on the partition function. Our work is based on the recent algorithm of Hazan and Jaakkola \cite{Hazan12}. 

\section{Related Work}
Methods to \emph{approximate} the partition function for interacting RNAs have \emph{not} been proposed in the literature. Instead, methods for \emph{exact} comutation of the partition function have been developed, having high both time and space complexity. Most notably, \cite{Chitsaz09} developed an $O(n^6)$--time and $O(n^4)$--space dynamic programming algorithm that computes the partition function of RNA--RNA interaction complexes, thereby providing detailed insights into their thermodynamic properties. 
\cite{fenix10} has developed a \emph{sampling} algorithm that produces a Boltzmann weighted ensemble of RNA–-RNA interaction structures for the calculation of \emph{interaction probabilities} (and not the partition function) for any given interval on the target RNAs. 

In the context of \emph{single} RNA secondary structure prediction, \cite{Feng10} devised a Metropolis Monte Carlo algorithm, called ``Wang and Landau'' algorithm \cite{WangLandau01}, to approximate the partition function as well as density of states. 
Although the computation of the partition function over
all secondary structures and over all pseudoknot-free hybridizations
can be done by efficient dynamic programming algorithms, the real advantage 
of \cite{Feng10} is in approximating the partition function where pseudoknotted structures
are allowed; a context known to be NP-complete \cite{Lyngso00}.

In the machine learning community, there has been extensive research on 
obtaining  \emph{non-deterministic} and \emph{deterministic} approximations 
of the log-partition function.
Firstly, sampling and Monte Carlo methods (e.g. Gibbs Sampling and Monte Carlo Markov Chain) have been developed as non-deterministic approaches for estimating the partition function (cf. \cite{Koller09} and references therein). In high dimensions, obtaining independent samples from a given distribution is difficult since the mixing time is typically exponential in the size of the problem. Therefore, these methods are computationally very demanding, and in practice, they are rarely applied to large-scale problems.
Secondly, variational techniques have been extensively developed 
as a deterministic approach to efficiently estimate the partition function in
large-scale problems. In this approach, a simpler distribution is optimized as
an approximation to the \emph{true} distribution in a KL-divergence
sense. However, the difficulty of this approach comes from: (i) Non-convexity of the
set of feasible distributions (e.g. in ``mean field'' approximation \cite{Jordan99}), and/or (ii) Hardness of computing the entropy embedded in the KL objective. 
Variational upper bounds on the other hand are convex, usually derived by replacing
the entropy term in the KL objective with a simpler surrogate function
and relaxing constraints on sufficient statistics hence convexifying the set of feasible distributions \cite{Wainwright05}.

The basis of our work is \cite{Hazan12} which provides a framework for approximating and bounding the partition function using MAP\footnote{Maximum a posteriori} inference (i.e. prediction of the most likely structure) on randomly perturbed models. 
Particularly, they propose to estimate the partition function as the max-statistics of collections
of random variables, which is a major topic in \emph{extereme value statistics} (e.g. see \cite{yuri04}). More broadly, there is an emerging body of work on perturbation methods, showing the benefits of explicitly adding noise into the modeling, learning, and inference pipelines \cite{Papandreou11a,Tarlow12}. 

\section{Preliminaries}\label{sec:pre}
\subsection{Notation}
The input nucleic acid sequences are denoted by $\RR$ and $\RS$ throughout this paper. Function $L$ denotes the length of the input sequence, and $\RR$ is indexed from $1$ to $L(R)$, and $\RS$ is indexed from $1$ to $L(S)$ both in $5'$ to $3'$ direction. We refer to the $i^{th}$ nucleotide in $\RR$ and $\RS$ by $i_R$ and $i_S$ respectively. The subsequence from the $i^{th}$ nucleotide to the $j^{th}$ nucleotide in a strand is denoted by $[i, j]$. An intramolecular base pair between the nucleotides $i$ and $j$ in a strand is called an {\it arc} and denoted by a bullet $i \bullet j$. 
An intermolecular base pair between the nucleotides $i_R$ and $i_S$ is called a {\it bond} and denoted by a circle $i_R \circ i_S$. We denote an RNA (RNA-RNA interaction) secondary structure by $\mathfrak{s}$, which is mathematically a set of constituent base pairs (arcs and bonds). The collection of all such feasible structures is denoted by $\mathfrak{S}$.

Throughout this paper, we denote the partition function by
\begin{equation}
Q := \sum_{\mathfrak{s} \in \mathfrak{S}} e^{-\frac{G({\mathfrak{s})}}{RT}},
\end{equation}
in which $G(\mathfrak{s})$ is the free energy of $\mathfrak{s}$, $R$ is the gas constant, and $T$ is temperature. For a more detailed presentation of the partition function see \cite{Mcc90,Chitsaz09}. We use the Turner energy model for single RNAs \cite{MatTur99} and our energy model for RNA-RNA interaction \cite{Chitsaz09}.

In this paper, we consider only the canonical base pairing system, i.e. each nucleotide is Watson-Crick paired with at most one nucleotide. We also assume there are no pseudoknots in individual secondary structures of $\RR$ and $\RS$, and there are no crossing bonds and zigzags between $\RR$ and $\RS$ \cite{Chitsaz09}. However, an extension of our ideas to non-canonical base pairing systems \cite{Siederdissen11} and pseudoknotted (crossing and zigzagged) structures \cite{Sheikh12} is straight forward. The key requirement for such an extension is the existence of a fast minimum free energy structure prediction algorithm  that can incorporate per base-pair energy contributions for the considered class of structures.

\subsection{Energy Perturbations}
Following the Hazan-Jaakkola's approach \cite{Hazan12}, let $\{\gamma_{i_R \bullet j_R}\}$, $\{\overline{\gamma}_{i_R \bullet j_R}\}$,$\{\gamma_{i_S \bullet j_S}\}$, $\{\overline{\gamma}_{i_S \bullet j_S}\}$, $\{\gamma_{i_R \circ i_S}\}$, and $\{\overline{\gamma}_{i_R \circ i_S}\}$ be six families of independent and identically distributed (i.i.d.) random variables, which are energy perturbations corresponding to presence and absence of base pairs, following the Gumbel distribution whose cumulative distribution function is
\begin{equation}\label{equ:gumbel}
F(x) := P[\gamma \leq x] = e^{-e^{-(x+C)}}.
\end{equation}
Above, $C$ is the Euler's constant, so that the mean of our Gumbel distribution defined in (\ref{equ:gumbel}) is zero.
For every structure $\mathfrak{s} \in \mathfrak{S}$, let the energy perturbation of a structure be
\begin{equation}\label{equ:perturbs}
\gamma(\mathfrak{s}) = \sum_{i \bullet j \in \mathfrak{s}} \gamma_{i \bullet j} + \sum_{i \bullet j \not \in \mathfrak{s}} \overline{\gamma}_{i \bullet j}
\end{equation}
if $\mathfrak{s}$ is single RNA structure and 
\begin{equation}\label{equ:perturbi}
\begin{split}
\gamma(\mathfrak{s}) & = \sum_{i_R \bullet j_R \in \mathfrak{s}} \gamma_{i_R \bullet j_R} + \sum_{i_R \bullet j_R \not \in \mathfrak{s}} \overline{\gamma}_{i_R \bullet j_R} + \sum_{i_S \bullet j_S \in \mathfrak{s}} \gamma_{i_S \bullet j_S} + \sum_{i_S \bullet j_S \not \in \mathfrak{s}} \overline{\gamma}_{i_S \bullet j_S} \\
& \quad + \sum_{i_R \circ i_S \in \mathfrak{s}} \gamma_{i_R \circ i_S} + \sum_{i_R \circ i_S \not \in \mathfrak{s}} \overline{\gamma}_{i_R \circ i_S}
\end{split}
\end{equation}
if $\mathfrak{s}$ is RNA-RNA interaction structure. 


\section{Upper Bound on the Partition Function}

Corollary 1 in \cite{Hazan12} states that
\begin{equation} \label{equ:ub}
\begin{split}
\log Q &\leq E_\gamma \left[ \max_{\mathfrak{s} \in \mathfrak{S}} \{ -G(\mathfrak{s})/RT + \gamma(\mathfrak{s}) \} \right] \\
&= - E_\gamma \left[ \min_{\mathfrak{s} \in \mathfrak{S}} \{ G(\mathfrak{s})/RT - \gamma(\mathfrak{s})\} \right],
\end{split}
\end{equation}
in which $E$ is the expectation with respect to $\gamma$'s. 
The perturbations $\gamma$ include terms that depend on base pairs ($\gamma_{\cdot}$) and terms that depend on lack of base pairs ($\overline{\gamma}_{\cdot}$). To simplify the incorporation of $\overline{\gamma}$ terms into the energy model, let
\begin{equation}\label{equ:prtbs}
\lambda(\mathfrak{s}) = \sum_{i \bullet j \in \mathfrak{s}} (\gamma_{i \bullet j} - \overline{\gamma}_{i \bullet j})
\end{equation}
for single RNA structure and 
\begin{equation}\label{equ:prtbi}
\begin{split}
\lambda(\mathfrak{s}) & = \sum_{i_R \bullet j_R \in \mathfrak{s}} (\gamma_{i_R \bullet j_R} - \overline{\gamma}_{i_R \bullet j_R}) + \sum_{i_S \bullet j_S \in \mathfrak{s}} (\gamma_{i_S \bullet j_S} - \overline{\gamma}_{i_S \bullet j_S}) \\
& \quad + \sum_{i_R \circ i_S \in \mathfrak{s}} (\gamma_{i_R \circ i_S} - \overline{\gamma}_{i_R \circ i_S})
\end{split}
\end{equation}
for RNA-RNA interaction structure. Since the difference between two random variables following the Gumbel distribution follows the logistic distribution, we can rewrite the single RNA $\lambda(\mathfrak{s})$ as
\begin{equation}\label{equ:lambda_prtbs}
\lambda(\mathfrak{s}) = \sum_{i \bullet j \in \mathfrak{s}} \lambda_{i \bullet j}
\end{equation}
and the RNA-RNA interaction one as
\begin{equation}\label{equ:lambda_prtbi}
\lambda(\mathfrak{s}) = \sum_{i_R \bullet j_R \in \mathfrak{s}} \lambda_{i_R \bullet j_R} + \sum_{i_S \bullet j_S \in \mathfrak{s}} \lambda_{i_S \bullet j_S} + \sum_{i_R \circ i_S \in \mathfrak{s}} \lambda_{i_R \circ i_S},
\end{equation}
where $\lambda$'s are independent identically distributed random variables following the logistic distribution. In that case, (\ref{equ:perturbs}) and (\ref{equ:prtbs}) imply
\begin{equation}
\gamma(\mathfrak{s}) = \sum_{i \bullet j} \overline{\gamma}_{i \bullet j} + \lambda(\mathfrak{s})
\end{equation}
and (\ref{equ:perturbi}) and (\ref{equ:prtbi}) imply
\begin{equation}
\gamma(\mathfrak{s}) = \sum_{i_R \circ i_S} \overline{\gamma}_{i_R \circ i_S} + \sum_{i_R \bullet j_R} \overline{\gamma}_{i_R \bullet j_R} + \sum_{i_S \bullet j_S} \overline{\gamma}_{i_S \bullet j_S} + \lambda(\mathfrak{s}). 
\end{equation}
Since $\sum_{i \bullet j} \overline{\gamma}_{i \bullet j}$ and $\sum_{i \circ j} \overline{\gamma}_{i \circ j}$ are constants inside the minimization in (\ref{equ:ub}), whose expectations are zero because of (\ref{equ:gumbel}), we can rewrite (\ref{equ:ub}) as follows where $\lambda$ follows the logistic distribution:
\begin{equation}\label{equ:lambda_ub}
\log Q \leq - E_\lambda \left[ \min_{\mathfrak{s} \in \mathfrak{S}} \{ G(\mathfrak{s})/RT - \lambda(\mathfrak{s})\} \right].
\end{equation}

Our algorithm computes the right hand side of (\ref{equ:lambda_ub}) and calculates the upper bound $Q_{ub} = \exp(- E_\lambda \left[ \min_{\mathfrak{s} \in \mathfrak{S}} \{ G(\mathfrak{s})/RT - \lambda(\mathfrak{s})\} \right])$. The minimization inside the expectation is essentially minimum free energy prediction, albeit with a perturbed energy. Recall that the energy perturbation $-\lambda(\mathfrak{s})$ is the sum of individual base-pair perturbations for all base-pairs in $\mathfrak{s}$; therefore, incorporation of such a perturbation in fast minimum free energy prediction algorithms, such as \cite{Backofen11} which exploits sparsity, is straight forward. Particularly, we only need to add $-\lambda_{i \bullet j}$ to the calculation of $L^c(i,j)$ in \cite{Backofen11} when $i$ and $j$ can form a base-pair. Additionally, the scaling of energy by $RT$ in (\ref{equ:lambda_ub}) has to be carefully applied to the sparse algorithm. Similarly, we only need to add $-\lambda_{i_R \circ i_S}$ to the calculation of hybrid components in \cite{Salari10}, in addition to proper handling of $-\lambda_{i_R \bullet j_R}$ and $-\lambda_{i_S \bullet j_S}$ in intramolecular base-pairings.

\subsection{Complexity Analysis}
Note that the perturbed energy is such that the triangle inequality in Property 1 of \cite{Backofen11} still holds. Therefore,  the running time of all sparse folding algorithms based on the triangle inequality \cite{Backofen11,Mohl10,Salari10} is not affected by our energy perturbation. To calculate the expectation above, we sample $\lambda$'s independently from the logistic distribution until the estimation of the 
expectation by simple averaging converges. Our experimental results show that the number of samples needed is much lower than the size of the input sequence. 

For a single RNA with length $n$, the time complexity of our upper bound algorithm is $O([n^2 + MFE(n)]\ell)$ in which $\ell \leq n$ is the number of samples needed for the expectation estimation to converge, and $MFE(n)$ is the running time of minimum free energy prediction. In this case, the space complexity is $O(n^2 + MFES(n))$, in which $MFES(n)$ is the memory space needed for minimum free energy prediction.

For RNA-RNA interaction with lengths $m$ and $n$, the time complexity of our algorithm is $O\left(\left[m^2 + n^2 + MFE(m, n)\right]\ell\right)$ in which $\ell \leq n + m$ is the number of samples needed for the expectation estimation to converge, and $MFE(m, n)$ is the running time of minimum free energy prediction. In this case, the space complexity is $O(m^2 + n^2 + MFES(m, n))$, in which $MFES(m, n)$ is the memory space needed for minimum free energy prediction. Usually the running time and space complexity of our upper bound are dominated by those of minimum free energy prediction; therefore in practice, the time complexity of our upper bound is $O(MFE(n)\ell)$ for single RNA and $O(MFE(m, n)\ell)$ for RNA-RNA interaction, and its space complexity is often the same as that of minimum free energy prediction.

\begin{figure}[ph!]
\begin{center}
\includegraphics[width=0.7\textwidth]{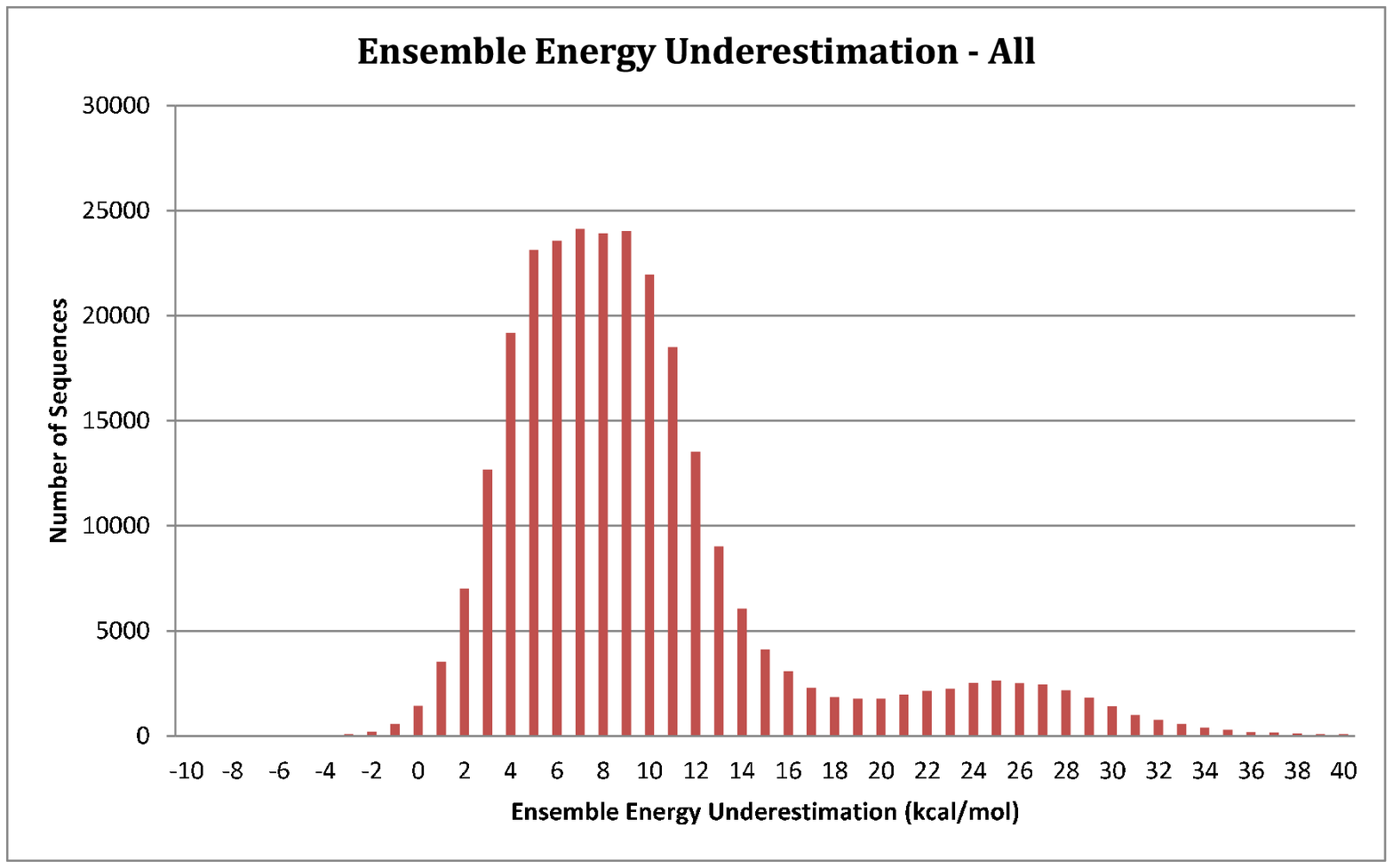}\\
\includegraphics[width=0.7\textwidth]{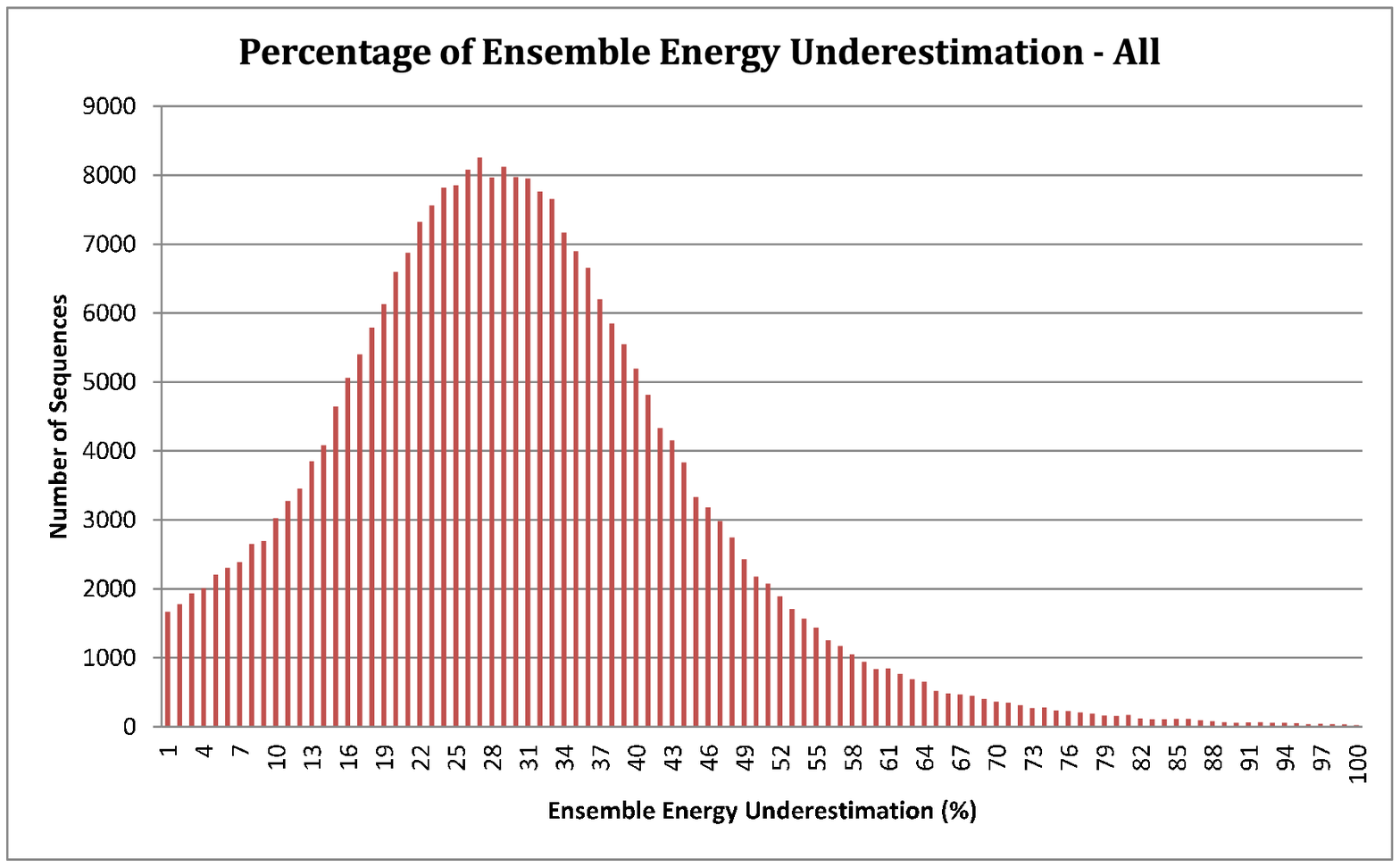}\\
\includegraphics[width=0.7\textwidth]{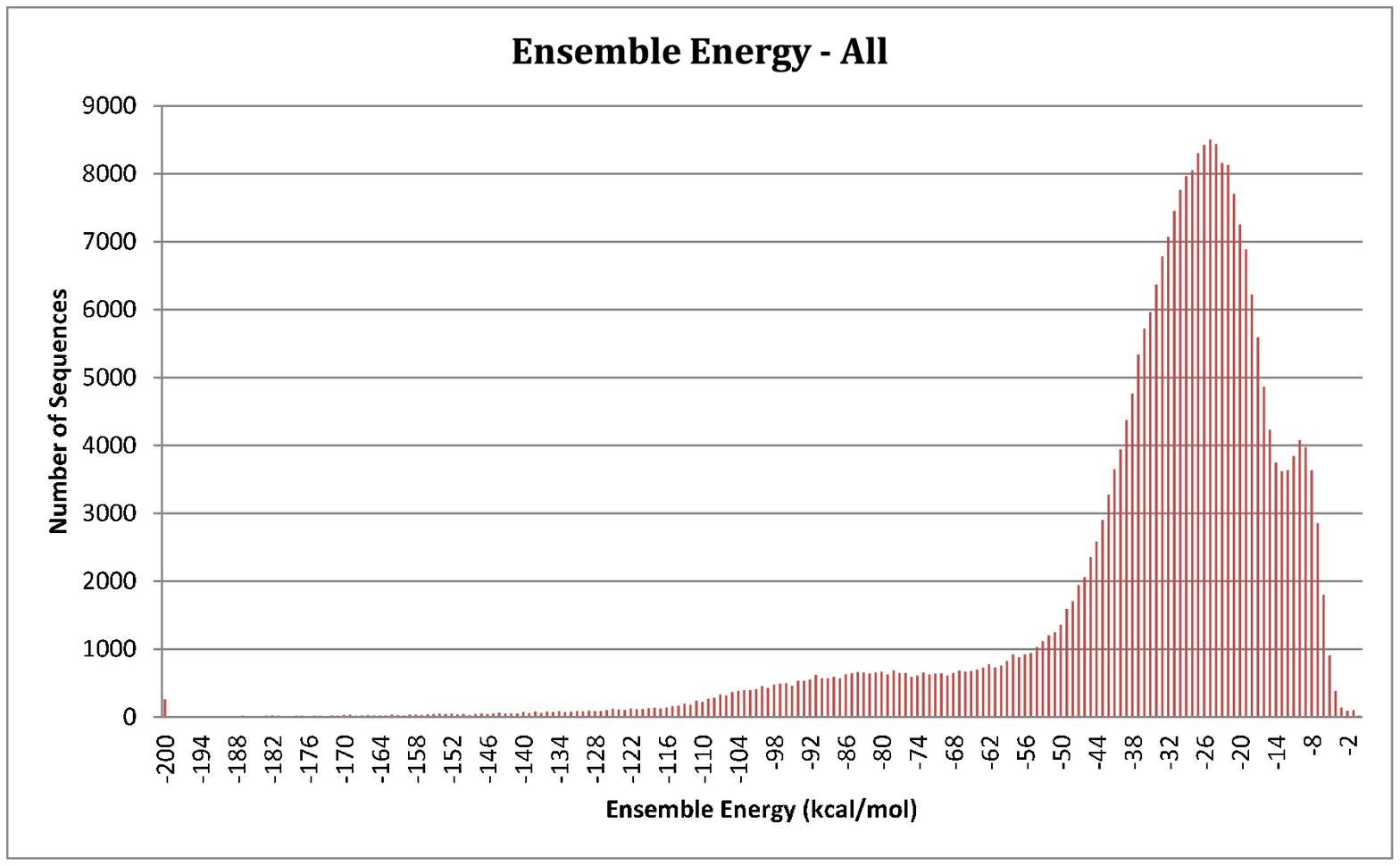}
\end{center}
\caption{Histograms of (top) ensemble energy underestimation $-RT\log Q + RT\log Q_{ub}$ for all 273,512 sequences in our dataset, (middle) percentage of ensemble energy underestimation $\log Q_{ub}/\log Q - 1$, and (bottom) ensemble energy $-RT\log Q$ in the input dataset.}\label{fig:underest}
\end{figure}
\begin{figure}[ph!]
\begin{center}
\includegraphics[width=0.7\textwidth]{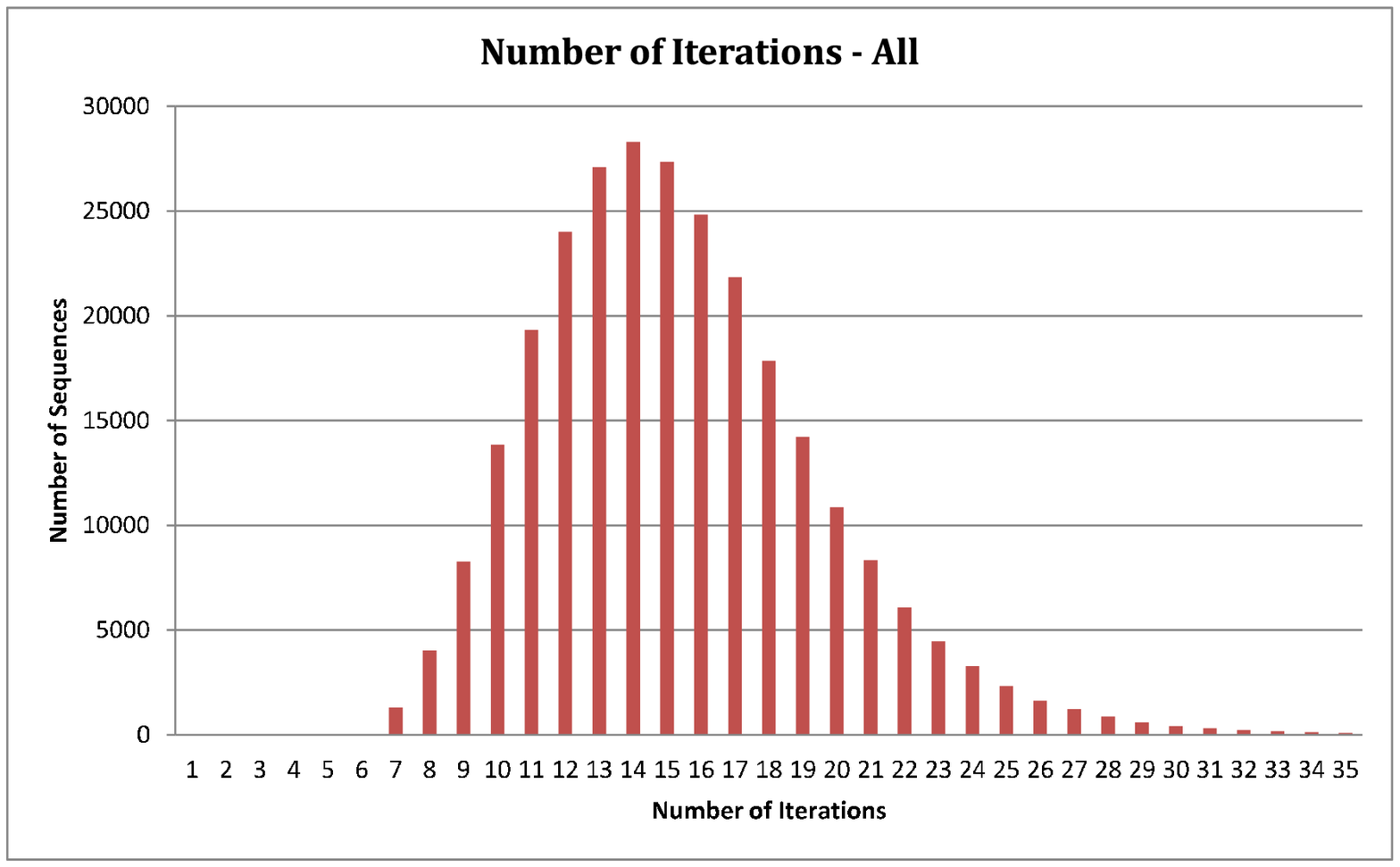} \\
\includegraphics[width=0.7\textwidth]{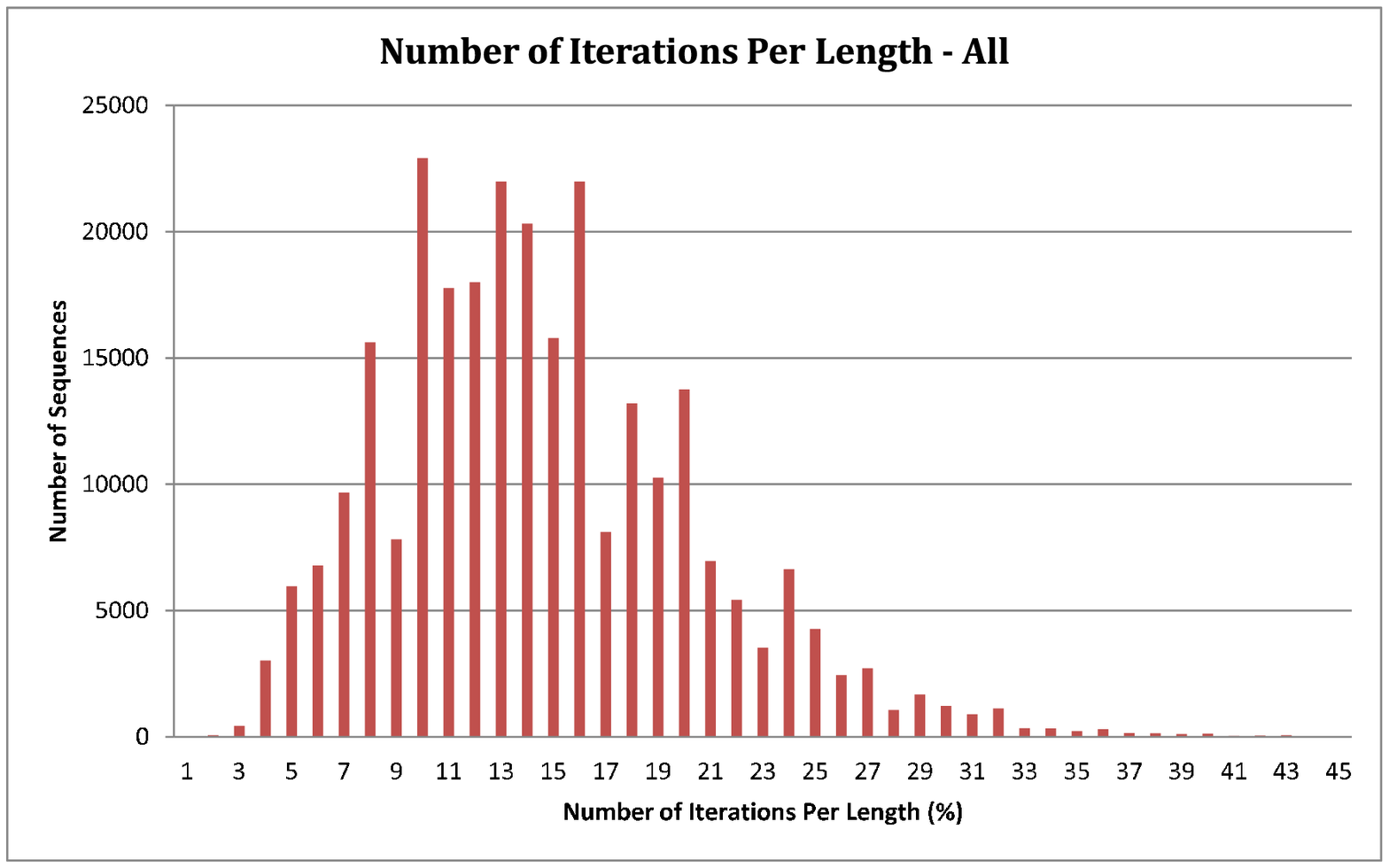} \\
\includegraphics[width=0.7\textwidth]{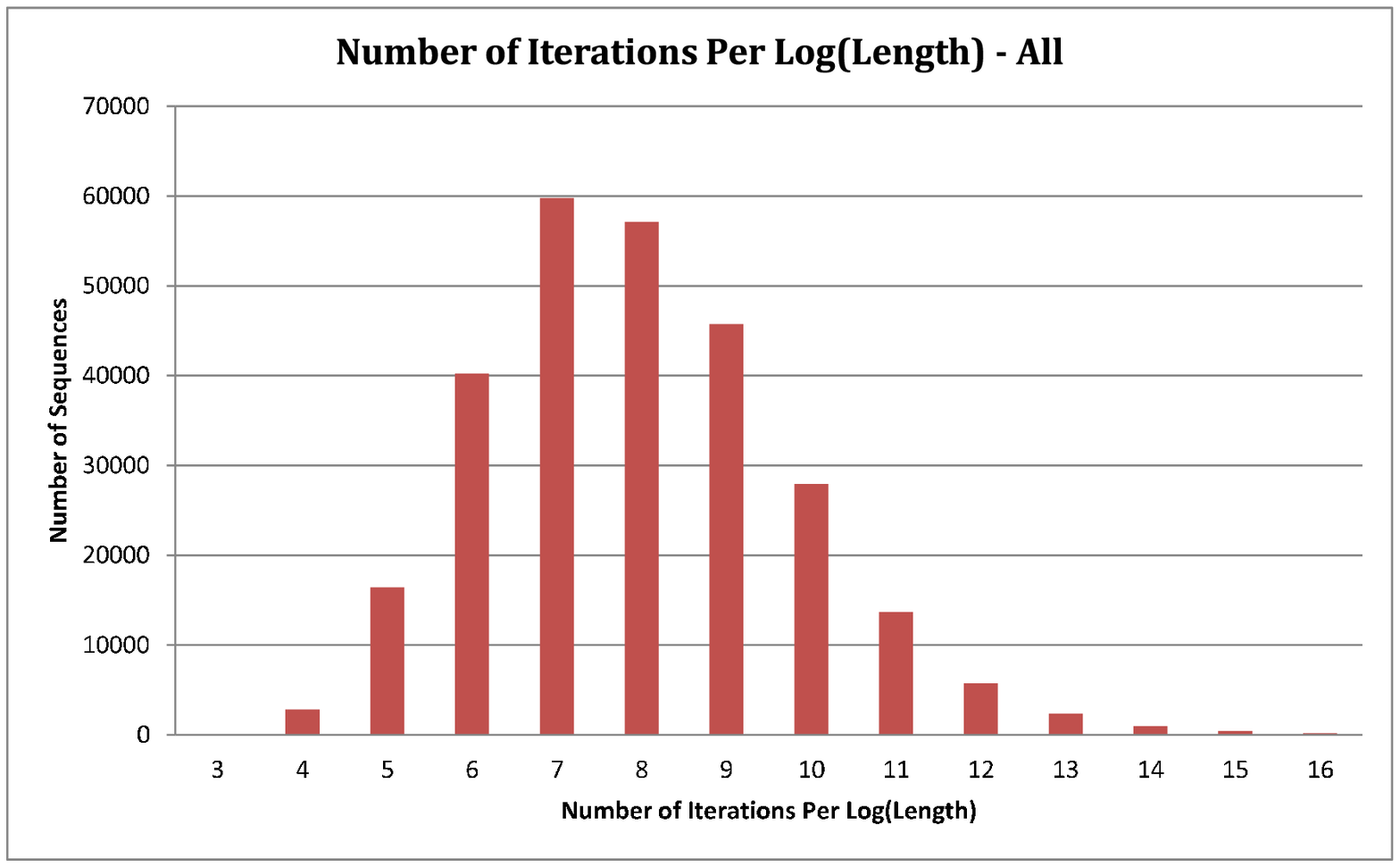}
\end{center}
\caption{Histograms of (top) number of iterations $\ell$ needed for all sequences in our dataset, (middle) number of iterations as a percentage of the input sequence length, and (bottom) number of iterations per the log of the input sequence length.}\label{fig:iters}
\end{figure}

\section{Results}
We implemented the upper bound algorithm in our {\tt piRNA} package \cite{Chitsaz09}. To test the overestimation of partition function and also the number of samples needed for the algorithm to converge, we randomly selected totally 273,512 RNA sequences from Rfam 11.0 database \cite{Rfam11,Rfam03} and computed the upper bounds for them. Since the partition functions are often large numbers, we report the ensemble energies. The ensemble energy is proportional to the minus log of the partition function and an overestimation of the partition function becomes an underestimation of the ensemble energy. Fig. \ref{fig:underest} depicts the results of our experiment. The top plot shows the histogram of ensemble energy underestimation calculated by $-RT\log Q + RT\log Q_{ub}$. The middle plot shows the histogram of ensemble energy underestimation percentage calculated from the ratio $\log Q_{ub} / \log Q - 1$. This plot shows that for a vast majority of cases this ratio is below $40\%$. The bottom plot depicts the distribution of ensemble energy $-RT \log Q$ in the input dataset. Although this distribution exhibits multiple peaks, the middle distribution, which is the underestimation percentage, has a unimodal behaviour. Out of 273,512 RNAs, for 249,622 of them the underestimation is less than $50\%$ of their ensemble energy, and for about half of the sequences (148,762) this difference is smaller than $30\%$. The number of sequences for which this difference is negative is 2,356 which is less than 1\% of the total number of RNAs. 

Fig. \ref{fig:iters} shows the performance of our algorithm. The top plot is the histogram of the number of samples $\ell$ (iterations) needed for the algorithm to converge. The middle and bottom plots show the histogram of the number of iterations per input size and per the log of input size. The vast majority of sequences required less than $15\%$ of their length iterations to converge. The number of iterations starts with 7 and almost all the sequences need less than 40 samples. 243,119 or $89\%$ of RNAs need not more than 20 iterations, and for more than half of the sequences (153,515) the experiment has been done with less than 15 iterations. Therefore at most 40 iterations are enough for different Rfam RNAs with different lengths. The number of iterations per length ratio for most of the sequences is less than $30\%$. For $90\%$ of the RNAs in this dataset, this number is between $7\%$ and $25\%$.  Clearly the relation between length and the number of iterations is not linear, and upper bounds for different RNAs with the same length require different number of iterations.

Recall that the space complexity of our algorithm is the same as that of sparse minimum free energy prediction. Therefore, our algorithm is both fast and space efficient in practice.


\section{Conclusion}
We gave a fast algorithm, based on the recent algorithm of Hazan and Jaakkola \cite{Hazan12}, to iteratively compute an upper bound on the partition function of nucleic acids by perturbing energy. Our upper bound algorithm uses a fast minimum free energy prediction in each iteration. Our algorithm preserves the properties on which sparsification methods rely; therefore, we minimally modified sparse folding algorithms \cite{Backofen11,Mohl10,Salari10} to obtain the required fast minimum free energy prediction.  

For the lower bound, one can trivially use the single term corresponding to the minimum free energy. The lower bound algorithm of Hazan and Jaakkola \cite{Hazan12} requires modification to be applicable to our problem.  We leave such more accurate lower bound algorithms for future work.


\bibliographystyle{splncs_srt}
\bibliography{pub,ref}

\end{document}